\documentclass[preprint,prd,amsmath,nofootinbib,twoside]{revtex4}
\usepackage{braket}
\usepackage{graphicx}
\usepackage{hyperref}

\begin{document}
\title{New Constraints on Neutrino Velocities}
\date{\today}

\author{Andrew G. Cohen}
\email{cohen@bu.edu}
\author{Sheldon L. Glashow}
\email{slg@bu.edu}
\affiliation{Physics Department,  Boston University\\
  Boston, MA 02215, USA}

\begin{abstract}
  The OPERA collaboration has claimed that muon neutrinos with mean
  energy of 17.5~GeV travel 730 km from CERN to the Gran Sasso at a
  speed exceeding that of light by about 7.5 km/s or 25~ppm.  However,
  we show that such superluminal neutrinos would lose energy rapidly
  via the bremsstrahlung of electron-positron pairs ($\nu\rightarrow
  \nu+e^-+e^+$). For the claimed superluminal neutrino velocity and at
  the stated mean neutrino energy, we find that most of the neutrinos
  would have suffered several pair emissions en route, causing the
  beam to be depleted of higher energy neutrinos.  Thus we refute the
  superluminal interpretation of the OPERA result.  Furthermore, we
  appeal to Super-Kamiokande and IceCube data to establish strong new
  limits on the superluminal propagation of high-energy neutrinos.
\end{abstract}

\maketitle{}

\section{Introduction and Conclusions}

The OPERA collaboration has reported evidence of superluminal neutrino
propagation\cite{Adams:2011zb}. The CNGS beam, consisting of pulses of muon
neutrinos with mean energy of 17.5~GeV and an energy spread extending
beyond 50~GeV, travels about 730~km from CERN to the OPERA detector in
the Gran Sasso Laboratory. The group reports that the travel time of
the ultrarelativistic neutrinos is about 60~ns less than expected.  We
phrase our discussion in terms of the parameter $\delta\equiv
(v_\nu^2-1)$ wherein we take the speed of light in vacua to be
unity. The OPERA claim is $\delta=5\times 10^{-5}$.  Recognizing the
potential impact of this result, the collaboration writes that it
intends ``to continue its studies to investigate
possible\ldots systematic effects that could explain the observed
anomaly.''  

The OPERA claim (hereafter, the anomaly) is compatible with earlier
studies of high-energy neutrinos such as MINOS\cite{Adamson:2007zzb},
which yielded the result $\delta =10.2 \pm 5.8 \times
10^{-5}$. However, observations of $\sim 10\ \text{MeV}$ neutrinos
from supernova SN1987a provide the
constraint\cite{Hirata:1987hu,Bionta:1987qt,Longo:1987ub} $\delta <
4\times 10^{-9}$. Thus, the alleged anomaly must be energy dependent,
decreasing rapidly from 10 GeV to 10 MeV.  We note in passing that
observations of neutrino oscillations allow one to deduce far more
severe constraints on neutrino velocities at relevant
energies\cite{Coleman:1998ti,Coleman:1997xq}. Lorentz-violating
velocity differences as large as $10^{-20}$ between neutrinos of
different species would have been readily detected and are
excluded. Thus, the velocity anomaly, if correct, must pertain to the
propagation of all three types of neutrino.

Let us assume that muon neutrinos with energies of order tens of GeV
travel at superluminal velocity. As in all cases of superluminal
propagation, certain otherwise forbidden processes are kinematically
permitted, even in vacuum. In particular, we focus on the following
analogs to Cherenkov radiation:
\begin{equation}
  \label{eq:1}
  \nu_\mu\longrightarrow \left\{
    \begin{aligned}
      &\nu_\mu+\gamma &(a)\\
      &\nu_\mu+ \nu_e + \overline{\nu}_e &(b)\\
      &\nu_\mu + e^+ + e^- &(c)
    \end{aligned}
    \right.
\end{equation}
These processes cause superluminal neutrinos to lose energy as they
propagate and, as we shall see, process (c) places a severe constraint
upon potentially superluminal neutrino velocities.

Process (b) is irrelevant because all three neutrino species are known
to travel at virtually the same velocity.  Process (a) is
kinematically allowed for all neutrino energies provided $v_{\nu} >
1$.  However process (a) is induced by a $W$ loop diagram and thus we
find its effect on neutrino propagation to be smaller by a factor of
$\alpha/\pi$ than that of process (c), whenever process (c) is
kinematically allowed.

Process (c), pair bremsstrahllung, proceeds through the neutral
current weak interaction.  The threshold energy for this process is
$E_0= 2m_e/\sqrt{v_\nu^2- v_e^2}$, where $v_e$ is the maximal
attainable velocity of an electron and $m_e$ its mass.  However, we
know\cite{Coleman:1998ti,Coleman:1997xq} that $v_e=1$ to a precision
of at least $10^{-15}$.  Thus we may write
$E_0=2m_e/\sqrt{\delta}$. Its value is about 140~MeV for the OPERA
value of $\delta$.  It is process (c) that allows us to exclude the
OPERA anomaly and place a strong constraint on neutrino
superluminality.

We have computed both $\Gamma$, the rate of pair emission by an
energetic superluminal neutrino, and $dE/dx$, the rate at which it
loses energy in the high energy limit where the electron and neutrino
masses may be neglected\footnote{These expressions are leading order
  in $\delta$. We have also neglected the vector-current coupling of the
  electron: $c_{V}=0$ and $c_{A}=-1/2$.}:
\begin{gather}
  \label{eq:3}
  \Gamma = k' \frac{G_{F}^{2} }{192\pi^3} E^{5}\delta^3\\
  \frac{dE}{dx} = -k \frac{G_{F}^{2}}{192\pi^3} E^{6}\delta^3
\end{gather}
where $k$ and $k'$ are numerical constants: $k=25/448$, $k'= 1/14$.
These expressions, aside from the numerical factors, follow from
simple arguments. The factors of $G_{F}$ arise from the low energy
form of the weak interactions while those of energy then follow from
dimensional analysis. The power of $\delta$ is related to the power of
energy: the 4-momentum of the superluminal neutrino is timelike
(relative to the speed of light) with a square of $\delta E^{2}$. We
may therefore work in the neutrino ``rest'' frame with an effective
``mass'' of $\sqrt{\delta} E$. In this frame the powers of $\delta$
follow the powers of $E^{2}$. The relativistic dilation factor
needed to boost back to the original frame is the ratio of
the original energy divided by the effective ``mass'',
$\gamma = 1/\sqrt{\delta}$. Applying the usual dilation factors to 
$\Gamma$ and $dE/dx$ gives our result.

Note that the mean fractional energy loss due to a single pair
emission is $E^{-1}(dE/dx)/\Gamma = k/k' \approx 0.78$: about
three-quarters of the neutrino energy is lost in each emission.

We integrate $dE/dx$ assuming $\delta$ not to vary significantly in
the relevant energy interval.  We find that neutrinos with initial
energy $E_0$, after traveling a distance $L$, will have energy $E$
as given by:
\begin{equation}
  \label{eq:4}
  E^{-5}-E_0^{-5} = 5 k \delta^{3} \frac{G_{F}^{2}}{192\pi^{3}}L\equiv E_{T}^{-5}
\end{equation}
The steeply falling (with energy) form of $dE/dx$ means that neutrinos
with initial energy greater than $E_{T}$ rapidly approach a terminal
energy, $E_{T}$, which is essentially independent of the initial
neutrino energy.  Adopting the OPERA result $\delta=5\times 10^{-5}$
and using the OPERA baseline of $730 \text{ km}$ we find a terminal
energy of about 12.5 GeV. Few, if any, neutrinos will reach the
detector with energies in excess of 12.5 GeV.  Thus the CNGS beam
would be profoundly depleted and spectrally distorted upon its arrival
at the Gran Sasso.  Using the expression for $\Gamma$ above we may
also establish that \textit{any} superluminal neutrino with the
velocity claimed by OPERA of \textit{any} specific initial energy much
greater than 12.5 GeV has a negligible probability of arriving at the
Gran Sasso without having lost most of its energy.  The observation of
neutrinos with energies in excess of 12.5~GeV cannot be reconciled
with the claimed superluminal neutrino velocity measurement.

Our analysis yields strong new constraints on superluminal neutrino
velocities.  Super-Kamiokande has carefully studied atmospheric neutrinos that
traverse the earth (upward-going in the detector) over an energy range
extending from 1 GeV to 1
TeV\cite{Ashie:2005ik,Desai:2007ra,Swanson:2006gm}. These upward
directed neutrinos, in traversing a distance of 10,000~km, would
experience a depletion and spectral distortion as we have described
above. The observation of such neutrinos with 1 TeV energy allows us
to conservatively deduce that $\delta < 1.4\times 10^{-8}$, similar to
but slightly weaker than the lower energy neutrino velocity constraint
deduced from SN1987a.

The IceCube collaboration has reported the observation of upward-going
showers with reconstructed shower energies above 16
TeV\cite{Abbasi:2011ui}. Using a neutrino energy of 16 TeV and a
minimum baseline of $500 \text{ km}$ (which would be appropriate for a
horizontal neutrino) we obtain a more stringent limit $\delta < 3.75
\times 10^{-10}$, superior to the SN1987a constraint by an order of
magnitude. Finally IceCube has also reported events with energies in
excess of $100$ TeV. Observation of neutrinos with this energy and a
baseline of at least $500$ km implies a limit of $\delta < 1.7 \times
10^{-11}$.

While $\delta < 1.7 \times 10^{-11}$ is significantly better than
previous bounds, a more careful analysis of the path-lengths and
energies of the highest energy events from Super-Kamiokande, IceCube
and other neutrino telescopes may enable an even stronger constraint.

\begin{acknowledgments}
  We thank Ed Kearns for assistance in interpreting the neutrino data.
  We thank the students of Kilachand Honors College of Boston
  University for participating in impromptu conversations about the
  OPERA result. This work was supported by the U.S.\ Department of
  Energy Office of Science.
\end{acknowledgments}

\bibliography{supernus}
\bibliographystyle{utphys}

\end{document}